# Enhanced Mode Selection Algorithm for H.264 encoder for Application in Low Computational power devices

Sourabh Rungta
CSE Department.
RCET,
Durg, India.
E-mail: sourabh@rungta.org

Neeta Tripathi
ECE Department.
RSRCET,
Durg, India.
E-mail: neeta_31dec@rediffmail.com

Kshitij Verma
ABV-IIITM,
Gwalior, India.
Email: vermaksh@gmail.com

Anupam Shukla
ICT Department.
ABV-IIITM,
Gwalior, India.
E-mail: dranupamshukla@gmail.com

*Abstract*— **The intent of the H.264/AVC project was to create a standard capable of providing good video quality at substantially lower bit rates than previous standards without increasing the complexity of design so much that it would be impractical or excessively expensive to implement. An additional goal was to provide enough flexibility to allow the standard to be applied to a wide variety of applications. To achieve better coding efficiency, H.264/AVC uses several techniques such as inter mode and intra mode prediction with variable size motion compensation, which adopts Rate Distortion Optimization (RDO). This increases the computational complexity of the encoder especially for devices with lower processing capabilities such as mobile and other handheld devices. In this paper, we propose an algorithm to reduce the number of mode and sub mode evaluations in inter mode prediction. Experimental results show that this fast intra mode selection algorithm can lessen about 75% encoding time with little loss of bit rate and visual quality.**

## I    INTRODUCTION

H.264 is the emerging video coding standard with enhanced compression performance when compared to other existing coding standards to achieve outstanding coding performance, H.264/AVC employs several powerful coding techniques such as 4x4 integer transform, inter-prediction with variable block-size motion compensation, motion vector of quarter-pel accuracy, in-loop deblocking filter, improved entropy coding such as *context-adaptive variable-length coding* (CAVLC) and *content-adaptive binary arithmetic coding* (CABAC), enhanced intra-prediction, multiple reference picture, and the forth. Due to this new features, encoder computational complexity is extremely increased compared to previous standards. This makes H.264/AVC difficult for applications with low computational capabilities (such as mobile devices). Thus until now, the reduction of its complexity is a challenging task in H.264/AVC.

As recent multimedia applications (using various types of networks) are growing rapidly, video compression requires higher performance as well as new features. H.264 emerged as the video coding standard with enhanced video compression performance when compared to other existing coding standards. It outperforms the existing standards typically by a factor of two. Its excellent performance is achieved at the expense of the heavy computational load in the encoder. H.264/AVC has gained more and more attention; mainly due to its high coding efficiency (the average bitrate saving up to 50% as compared to H.263+ and MPEG-4 Simple Profile), minor increase in decoder complexity compared to existing standards, adaptation to delay constraints (the low delay mode), error robustness, and network friendliness.

H.264/AVC employs several powerful coding techniques such as 4x4 integer transform, inter-prediction with variable block-size motion compensation, motion vector of quarter-pel accuracy, in-loop deblocking filter, improved entropy coding such as *context-adaptive variable-length coding* (CAVLC) and *content-adaptive binary arithmetic coding* (CABAC), enhanced intra-prediction, multiple reference picture, and the forth. Note that DCT coefficients of intra-frames are transformed from intra prediction residuals instead of transforming directly from original image content. Especially, for the inter-frame prediction, H.264 allows blocks of variable size seven modes of different sizes in all, which are 16x16, 16x8, 8x16, 8x8, 8x4, 4x8 and 4x4, that can be used in inter-frame motion estimation/compensation. These different block sizes actually form a one or two level hierarchy inside a macroblock are supported along with the SKIP mode [1], as shown in Figure 1. Hence the computational complexity of motion estimation increases considerably as compared with previous standards. This is one major bottleneck for the H.264 encoder.

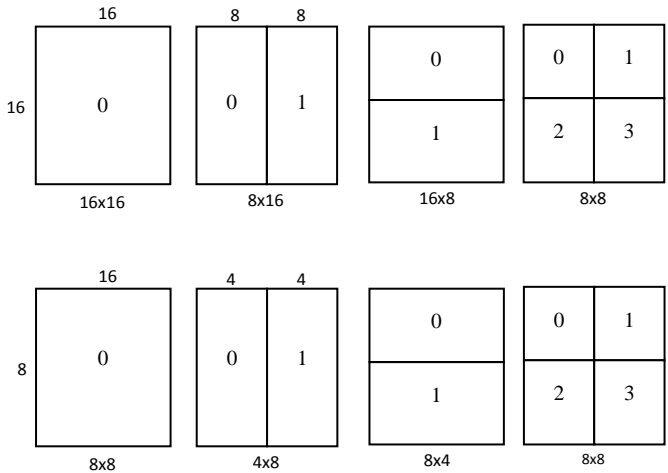

Figure 1.Macroblocks and Sub-macroblock partitions.

H.264 supports various intra-mode and inter-mode prediction techniques among which most of them contribute to the coding efficiency. Lagrangian RDO method is used to select the best coding mode of intra and inter prediction with highest coding efficiency [4]. In Inter prediction tree-structured multi-block sizes i.e. seven modes with different block sizes is supported by this standard. H.264 tests the encoding process with all possible coding modes of inter-coding, and calculates their RD costs to choose the mode having the minimum cost. RDO technique involves a lot of computations. The reference implementation [7] of H.264 uses a brute force search for inter mode selection which is extremely computational constraining. Therefore there is an obvious need for reducing the amount of modes that are evaluated in order to speed up the encoding and hence to reduce the complexity of the encoder.

## II   INTRA- AND INTER-FRAME SELECTION

The two coding modes in H.264 are intra-frame coding and inter-frame coding. Intra-frame coding supports two classes which are denoted as Intra4x4 and Intral6x16. When the subsequent frames of the video sequence have comparably large difference among them (such as in case of scene change), Intra-frame coding [1] would be selected in order to achieve outstanding coding performance, many advanced techniques are used. In these techniques, intra mode plays a vital role because it can eliminate spatial redundancy remarkably. In luma component, intra prediction is applied for each $4\times4$ block and for a $16\times16$ macroblock. There are 9 modes for $4\times4$ luma block, 4 modes for $16\times16$ luma block and 4 modes for $8\times8$ chroma block. In order to obtain the best coding performance, a very time-consuming technique named RDO (rate distortion optimization) is used. It computes the real bit-rate and distortion between original and reconstructed frames for each mode. Then it calculates the RDcost based on Lagrangian rate distortion formula. The mode which has the minimum RD cost will be chosen as the final coding mode. Therefore, the computational load of this kind of exhausting searching algorithm is not acceptable for real-time applications.

Inter prediction uses block-based motion compensation and it creates a prediction model from one or more previously encoded video frames or fields used. Encoding a motion vector for each partition can cost a significant number of bits, especially if small partition sizes are chosen. Motion vectors for neighboring partitions are often highly correlated and so each motion vector is predicted from vectors of nearby, previously coded partitions. A predicted vector, MVp, is formed based on previously calculated motion vectors and MVD, the difference between the current vector and the predicted vector, is encoded and transmitted. The method of forming the prediction MVp depends on the motion compensation partition size and on the availability of nearby vectors. H.264 supports a range of block sizes (from $16\times16$ down to 4x4) and fine subsample motion vectors (quartersample resolution) which are not supported by earlier standards. Inter-frame selection supports the following modes: SKIP, 16x16, 16x8, 8x16, 8x8, 8x4, 4x8 and 4x4. The mode decision is made by choosing the mode having minimum RDO cost [2].

$$J(s, c, MODE|\lambda_{MODE}) = SSD(s, c, MODE|QP) + \lambda_{MODE\ x}\ R(s, c, MODE|QP).$$

Where $J(s, c, MODE|\lambda_{MODE})$ represents the mode cost, QP denotes Quantization Parameter, $\lambda_{MODE}$ is the Lagrange multiplier [4] for mode decision, MODE indicates a mode chosen from the set of potential macroblock modes: {SKIP, 16 X 16, 16 X 8, 8X16, 8X8, 8X4, 4X8, 4X4, Intra_4 X 4,

Intra 16 X 16}, SSD represents the Sum of Squared Differences between the original block s and its reconstruction c.

SSD(s, c, MODE|QP) =

$$\sum_{x=1,y=1}^{16,16}(Sy[x,y]-Cy[x,y,MODE|QP])\,2$$

$$+\sum_{x=1,y=1}^{8,8}(Su[x,y]-Cu[x,y,MODE|QP])\,2$$

$$+\sum_{x=1,y=1}^{8,8}(Sv[x,y]-Cv[x,y,MODE|QP])\,2$$

where Cy[x,y,MODE|QP]) and Sy[x,y] represent the reconstructed and original luminance values; Cu, Cv and Su,Sv indicate the corresponding chrominance values, and R(x,y,MODE|QP) is the number of bits associated with choosing MODE, including the bits for the macro-block header, the motion, and all DCT coefficients.

## III    THE PROPOSED IMPROVEMENT OF FAST INTERMODE SELECTION ALGORITHM FOR H.264

The proposed algorithm, as shown in Figure 2, first checks the skip condition and then makes the decision between the Class 16 and Class 8 modes based on the factors - homogeneity and temporal movement [8]. Once the class is decided, with in the class then it uses sub-mode selection algorithm [7] to decide the best mode among the sub-modes.

Decision I: Compute the MB difference Δ for the current macro block. If Δ is very large (Δ>$\varphi_{inter}$) then intra mode selection is preferred.

Decision II: In this decision we first check the condition for the SKIP mode. If the current 16x16 block has no movement, i.e.Δ = 0 or Δ≤$\varphi$SKIP then SKIP mode is the best mode.

Decision III: Once SKIP is ruled out, we make a decision between Class 16 and Class 8 modes. Here we check the homogeneity of the block. If the macro block is homogeneous then Class 16 is chosen else Class 8 is chosen. The homogeneity of the macro block is determined by the Probability Based Macroblock Mode Selection.
Let P denote the probability of the current MB, then we have

$$P(x)\sum_{i=A,B,C,D}W_i\,.\,P_i(x)$$

A $Pmode$ cut set with $Pmode \,\epsilon\, [0,1]$ is used to determine the category which current MB belongs. Because we can get the probability of all modes, which are computed

dynamically every frame, we let this cut set equal to the probability of LMB. So we have

$$P_{mode} = 1 - \{P_{SKIP} + P_{INTER\,16\times16} + P_{INTER\,8\times8}\}$$

Correct Ratio is the probability of the MB classification to predict the same optimal encoding mode obtained from exhaustive mode selection, HMBErrRatio reflects the probability for HMBs to be mistakenly categorized as LMBs, while the LMBErrRatio reflects the probability for LMBs to be mistakenly categorized as HMBs. Compared with the classification accurate ratio of FMMS in, our algorithm shows the robust over all kinds of sequences with different motion and other features.

Decision IV: A MB is determined to be an LMB when the weighted sum is lower than $Pmode$, and if the $P_{mode}$ is higher than the minimum of $P_{SKIP} + P_{INTER\,16\times16} + P_{INTER\,8\times8}$, the MB is determined to be a true HMB. Otherwise, we need to further classify its motion character. Here a motion classifier is continuing used to determine if the MB contains complex motion information or simple motion information. By combining two types of classifiers, each MB can be efficiently categorized to different mode and motion search paths, which significantly reduces encoder complexity of H.264 for all types of content. Our fast mode decision algorithm consists of the following steps:

Step1: If the MB is in the first row or column of a frame, test all possible modes, select the best one, then exit.

Step2: Each MB is categorized by a probability classifier. If the predict mode is included in the HMBs, go to Step 4. Otherwise, go to Step 3.

Step3: Check the mode of INTER8 × 16 and INTER16 × 8. Go to Step 9.

Step4: For B picture, calculate the RD cost of direct mode. If it is lower than the threshold, which is defined as the minimum of neighboring MBs, skip all other modes and go to step 11. Otherwise, If the predict mode is included in the TRUE HMBs, go to Step 10, otherwise go to Step 5.

Step5: To categorize the MB with a motion classifier. If it has complex motion content, go to step 6. Otherwise, go to Step 8.

Step6: Check mode INTER8 × 8, INTER8 × 4, INTER4 × 8, INTER4 × 4. If there are more than two sub-macroblock modes are not INTER8 × 8, go to step 9. Otherwise, go to Step 7.

Step7: Check mode INTER16×16, INTER16×8 and INTER8×16. If any mode cost is more than INTER8×8 or the three modes have been tried, go Step 11.

Step8: Check mode INTER16×16 and INTER16×8, if cost16×16 < cost16×, go to Step 9. Otherwise, check all the other Inter modes.

Step9: Check INTRA16 × 16 and INTRA4 × 4.

Step10: Check INTER16 × 16 and INTER8 × 8.

Step11: Record the best MB mode and the minimum RD cost.

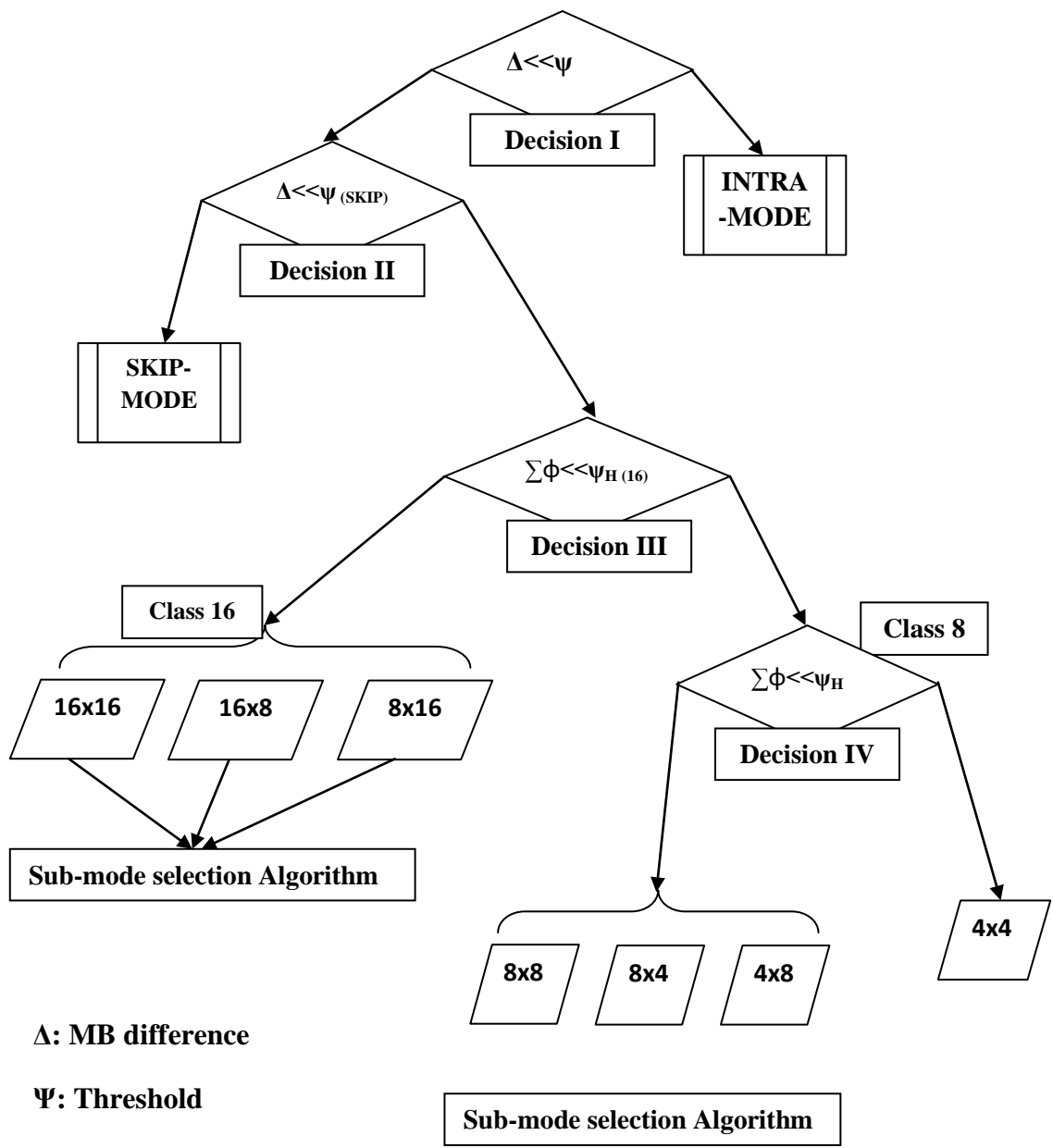

Figure 2: Decision Diagram (If the decision is yes, move to the left branch, else move to the right branch)

## IV    EXPECTED RESULTS

H.264 reference software JM8.6 [7] is applied as platform for evaluating the performance of the improved algorithm. We selected four representative QCIF video sequences i.e. Container, Foreman, Salesman and Coastguard as our test sequences.

TABLE I. SIMULATION PARAMETERS

| MV Search Range | 16 |
|---|---|
| GOP | IPPP |
| Codec | JM 8.6 |
| QP | 28, 32, 36, 40 |
| ProfileIDC | 66, 30 |
| Hadamard Transform | Used |
| Entropy Coding Method | CAVLC |
| Size | QCIF |
| Threshold set for Homogenity | 16×16: 20000 8×8: 5000 |

TABLE II.SIMULATION RESULTS FOR IPPP TYPE SEQUENCES

| Video Sequences | ΔTime (%) | ΔPSNR (dB) | ΔRate (%) |
|---|---|---|---|
| container_qcif.yuv | -86.69 | -0.04 | 0.40 |
| salesman_qcif.yuv | -77.16 | -0.03 | 0.91 |
| foreman_qcif.yuv | -69.50 | -0.10 | 1.38 |
| coastguard_qcif.yuv | -62.63 | -0.07 | 1.22 |

The test conditions [12] are shown in Table1.We used four Quantization Parameters while conducting the experiments on the test sequences, i.e. QP = 28, QP = 32, QP = 36 and QP = 40.

The coding parameters used to evaluate the efficiency are $\Delta T$, change of average PSNR – $\Delta PSNR$ and change of average date bits - $\Delta Bitrate$.

$T_{ref}$ is the coding time used by JM8.6 encoder. Let $T_{proposed}$ be the time taken by the proposed algorithm.
The $\Delta T$ % is defined as

$$\Delta T\% = \left[ \frac{T_{proposed} - T_{ref}}{T_{ref}} \right] \times 100$$

$$\Delta Bitrate\% = \left[ \frac{Bitrate_{proposed} - Bitrate_{ref}}{Bitrate_{ref}} \right] \times 100$$

The experimental results are shown in the Table 1. From Table 1, it is inevitable that the proposed algorithm reduces the encoding time for the four test sequences. Compared with the coding time of JM8.6 encoder, the coding time reduces by (88.92) % for slow motion videos, where as, it reduces by (70.1) % for fast motion videos. The PSNR degradation is up to (0.04 db) which is invisible to human eye and the data bits are increased up to (0.93) %.

## V    CONCLUSION

In this paper we proposed a fast inter mode selection algorithm based on the homogenous and temporal stationary characteristics of the video object and a procedure to select best submode. Verified by the fast, mild and slow motion sequences, our method could reduce the computational complexity of the H.264/AVC encoder by choosing the best mode judiciously. Average time reduction is about 75% in IPPP sequence. Moreover, our algorithm can maintain the video quality without significant bit-rate loss. It is helpful for the real-time implementation of the H.264 encoder and useful for the low-power applications of video coding.

## VI    REFERENCES

[1]. ThomasWiegand, Gary J. Sullivan, Senior Member, IEEE, Gisle Bjøntegaard, and Ajay Luthra, Senior Member, IEEE. Overview of the H.264/AVC Video Coding Standard.

[2]. Yun Cheng, Kui Dai, Jianjun Guo, Zhiying Wang, Minlian Xiao. Research on Intra Modes for Inter-Frame Coding in H.264 presented The 9th International Conference on Computer Supported Cooperative Work in Design Proceedings.

[3]. Iain E.G. Richardson. H.264 and MPEG-4 Video Compression, Wiley2004.

[4]. Jeyun Lee and Byeungwoo Jeon. Fast Mode Decision for H.264 with Variable Motion Block Sizes. Springer - Verlag, LNCS 2869, pages 723-730, 2003.

[5]. Iain E.G. Richardson. Video Codec Design, Wiley 2002.

[6]. Zhi Zhou and Ming-Ting Sun. Fast Macroblock Inter Mode Decisionand Motion Estimation for

H.264/MPEG- 4 AVC. Proceedings of IEEEInternational Conference on Image Processing, ICIP 2004, Singapore,pages 789-792, 2004.

[7]. H.264 Reference Software Version JM6.1d, http://bs.hhi.de/~ suehring/tml/, March 2003.

[8]. Mehdi Jafari, Islamic Azad University, Shohreh Kasaei, Sharif University of Technology. Fast Intra- and Inter- Prediction Mode Decision in H.264 Advanced Video Coding.

[9]. D. Wu, S. Wu, K. P. Lim, F. Pan, Z. G. Li, X. Lin. Block Inter Mode Decision for Fast Encoding of H.264. Institute for Infocomm Research (I2R) Agency for Science Technology and Research (A*STAR).

[10]. Iain Richardson and Yafan Zhao. Video Encoder Complexity Reduction by Estimation Skip Mode Distortion. Proceedings of IEEE International Conference on Image Processing, ICIP 2004, Singapore, pages 103- 106, 2004.

[11]. Keman Yu, Jiangbo Lv, Jiang Li and Shipeng Li. Practical Real-TimeVideo Codec for Mobile Devices. Proceedings of 2003 IEEE InternationalConference on Multimedia and Expo, ICME 2003, USA, pages 509-512, 2003.

[12]. Gary Sullivan, "Recommended Simulation Common Conditions for H.26L Coding Efficiency Experiments on Low Resolution Progressive Scan Source Material," VCEG-N81, 14th meeting: Santa Barbara, USA. Sept. 2001.

[13]. Iain Richardson H.264 and MPEG-4 Video Compression Video Coding for Next-generation Multimedia.

[14]. ISO/IEC 14496-10 and ITU-T Rec. H.264, Advanced Video Coding, 2003.

[15]. A. Hallapuro, M. Karczewicz and H. Malvar, Low Complexity Transform and Quantization – Part I: Basic Implementation, JVT document JVT-B038, Geneva, February 2002.

[16]. Zhenyu Wei, Hongliang Li and King Ngi Ngan, An Efficient Intra Mode Selection Algorithm For H.264 Based On Fast Edge Classification. Proceedings of 2007 IEEE International Symposium on Circuits and Systems, 2007, ISCAS 2007, New Orleans, LA, pages 3630-3633, 2007.

AUTHOR'S PROFILE

Anupam Shukla was born on 1st January 1965, at Bhilai (CG). He is presently working as an Associate Professor (Information Communication & Technology Deptt) at Atal Bihari Vajpayee Indian Institute of Information Technology & Management,(ABVIIITM), Gwalior (MP). He completed PhD (Electronics & Telecommunication) in the area of Artificial Neural Networks in the year 2002 and ME (Electronics & Telecommunication) Specialization in Computer Engineering in the year 1998 from Jadavpur University, Kolkata. He stood first position in the niversity and was awarded with gold medal. He completed BE (Hons) in Electronics Engineering in 1988 from MREC, JaipurHe has teaching experience of 19 years. His research area includes Speech recognition, Artificial neural etworks, Image Processing & Robotics. He published around 57 papers in national/international journals and conferences.

Sourabh Rungta is presently working as an Reader (Computer Science and Engineering Departtment) in RCET, Durg (CG). He completed M.Tech (Hons) in 2004. He completed BE in 1998. He has teaching experience of 5 years. He published around 5 papers in ational/international conferences and journals.

Neeta Tripathi is principle of RCET, Durg. She has eaching experience of 20 years. She published around 30 papers in National/international conferences and journals. Her contributed research area includes speech recognition.

Kshitij Verma is presently pursuing M.E.in VLSI Design from SSCET,Bhilai(C.G.) He completed BE in Electronics And Telecommunication in 2005 from RCET,Bhilai(C.G.).